\documentclass[twocolumn,prb,showpacs,preprintnumbers,amsmath,amssymb]{revtex4}
\usepackage{graphicx}
\usepackage{dcolumn}
\usepackage{bm}

\begin{document}

\title{Off-Fermi surface cancellation effects in spin-Hall
conductivity of a two-dimensional Rashba electron gas}
\author{C. Grimaldi$^{1,2}$}
\author{E. Cappelluti$^{3,4}$}
\author{F. Marsiglio$^{2,5}$}

\affiliation{$^1$ LPM, Ecole Polytechnique F\'ed\'erale de
Lausanne, Station 17, CH-1015 Lausanne, Switzerland}

\affiliation{$^2$DPMC, Universit\'e de Gen\`{e}ve, 24 Quai
Ernest-Ansermet, CH-1211 Gen\`{e}ve 4, Switzerland}

\affiliation{$^3$Istituto dei Sistemi Complessi, CNR-INFM, via dei
Taurini 19, 00185 Roma, Italy}

\affiliation{$^4$Dipartimento di Fisica, Universit\`a ``La
Sapienza'', Piazzale Aldo Moro 2, 00185 Roma, Italy}

\affiliation{$^5$Department of Physics, University of Alberta,
Edmonton, Alberta, Canada, T6G 2J1}

\begin{abstract}
We calculate the spin-Hall conductivity of a disordered
two-dimensional Rashba electron gas within the self-consistent
Born approximation and for arbitrary values of the electron
density, parametrized by the ratio $E_F /E_0$, where $E_F$ is the
Fermi level and $E_0$ is the spin-orbit energy. We confirm
earlier results indicating that in the limit $E_F/E_0 \gg 1$ the
vertex corrections suppress the spin-Hall conductivity. However,
for sufficiently low electron density such that $E_F\lesssim
E_0$, we find that the vertex corrections no longer cancel the
contribution arising from the Fermi surface, and they cannot
therefore suppress the spin current. This is instead achieved by
contributions away from the Fermi surface, disregarded in earlier
studies, which become large when $E_F\lesssim E_0$.
\end{abstract}

\maketitle

The spin-Hall effect, i.e., the generation of a spin polarized
current transverse to the direction of an applied external
electric field, has recently raised considerable interest in view
of its possible application in spintronics. Alongside the
extrinsic spin-Hall effect,\cite{dyakonov} generated by the
spin-orbit (SO) coupling to impurities and defects, much
theoretical effort has been devoted to the intrinsic spin-Hall
effect arising from the one-particle band structure of spin-orbit
coupled systems.\cite{zhang,sinova} The initial claim that for a
two-dimensional (2D) electron gas subject to the Rashba SO
coupling the intrinsic spin-Hall conductivity, $\sigma_{\rm sH}$,
is a universal constant ($\sigma_{\rm sH}=|e|/8\pi$,\cite{sinova}
where $e$ is the electron charge) has been shown to be invalid
even for an arbitrarily small concentration of impurities, which
reduce $\sigma_{\rm sH}$ to
zero.\cite{raimo,inoue,halpe,olga,loss} At the same time,
however, for other models of SO coupling, like, for example, the
three-dimensional (3D) Dresselhaus model,\cite{mal} the Luttinger
model for valence band holes,\cite{mura} or generalized Rashba
models taking into account non linear momentum dependences of the
SO interaction\cite{nomura,shytov,khaetskii} or a non-quadratic
unperturbed band spectrum,\cite{dassarma} $\sigma_{\rm sH}$ has
been found to be robust against non-magnetic impurity
scatterings, suggesting that the vanishing of $\sigma_{\rm sH}$
is a peculiar feature of the linear Rashba model. This is also
supported by rather general arguments which do not rely on the
specific scattering process.\cite{olga,rashba}

Within the linear response theory, the vanishing of $\sigma_{\rm
sH}$ in the Rashba model has been ascribed to a cancellation
effect of the spin-dependent part of the ladder current vertex in
the Born approximation for impurity
scattering.\cite{raimo,inoue,olga,loss} This cancellation
basically follows from the fact that, as long as the Fermi energy
$E_F$ is much larger than the spin-orbit energy $E_0$, the factor
$\tau^{-1}$ associated with the current vertex (where $\tau$ is
the electron life-time due to impurities) is balanced by the
factor $\tau$ arising from the product of two Greens functions
appearing in the current vertex kernel. However, similarly to
what happens for other properties (e.g., the conductivity in
impure metals), such kinds of balance effects are usually
peculiar to the assumption that $E_F$ is the largest energy scale
of the problem, and one may wonder if the cancellation mechanism
based on the vertex function described in
Refs.[\onlinecite{raimo,inoue,olga,loss}] is still valid when
$E_F$ is comparable with $E_0$. The clarification of this issue
is important not only to assess the role of vertex corrections in
a general context, but it is also quite crucial in view of the
recent progress made in fabricating systems with large SO
couplings for which the $E_F\gg E_0$ approximation may not be
appropriate. Examples of such large SO systems are, among others,
HgTe quantum wells,\cite{gui} the surface states of metals and
semimetals,\cite{rote,koro} and even the heavy Fermion
superconductor CePt$_3$Si.\cite{samokhin} However the most
striking example is provided by bismuth/silver(111) surface alloys
displaying quadratic unperturbed bands split by a Rashba energy
of about $E_0=220$ meV.\cite{grioni1} In this system the Fermi
energy can be tuned by doping with lead atoms in such a way that
$E_F$ may be larger or lower than $E_0$.\cite{grioni1,grioni2}

In this paper we investigate the spin-Hall conductivity in the
Born approximation of impurity scattering and for arbitrary
values of $E_F/E_0$. We find that, apart from the high density
limit $E_F/E_0\rightarrow \infty$, the spin-dependent part of the
vertex function is generally not zero, and increases as $E_F/E_0$
decreases, eventually reaching unity as $E_F\rightarrow 0$. In
this situation, the spin-Hall conductivity $\sigma_{\rm sH}$
would be nonzero if calculated along the lines of
Refs.[\onlinecite{raimo,inoue,olga,loss}], in contradiction with
the general arguments of Refs.[\onlinecite{olga,rashba}]. We
resolve this inconsistency by showing that $\sigma_{\rm sH}$ is
actually canceled by the contributions away from the Fermi
surface, which have magnitude equal to those on the Fermi
surface, but opposite in sign.

We consider a 2D Rashba electron gas whose Hamiltonian is
\begin{equation}
\label{ham1} H=\frac{\hbar^2
k^2}{2m}+\gamma(k_x\sigma_y-k_y\sigma_x),
\end{equation}
where $m$ is the electron mass, $\mathbf{k}$ is the electron
wave-number, and $\gamma$ is the SO coupling. The corresponding
electron dispersion consists of two branches
$E_k^s=\hbar^2(k+sk_0)^2/2m$ where $s=\pm 1$ is the helicity
number and $k_0=m\gamma/\hbar^2$. In the following, we
parametrize the SO interaction by the Rashba energy
$E_0=\hbar^2k_0^2/2m$ that, for a clean system, corresponds the
the minimum inter-band excitation energy for an electron at the
bottom of the lower ($s=-1$) band. For simplicity, we consider a
short-ranged impurity potential of the form $V(\mathbf{r})=V_{\rm
imp}\sum_i \delta(\mathbf{r}-\mathbf{R}_i)$, where the summation
is performed over random positions $\mathbf{R}_i$ of the impurity
scatterers with density $n_i$. The corresponding electron Green's
function is a $2\times 2$ matrix in the spin space,
\begin{equation}\label{green1}
G(\mathbf{k},i\omega_n)=\frac{1}{2}\sum_{s=\pm}
[1+s(\hat{k}_x\sigma_y-\hat{k}_y\sigma_x)]G_s(k,i\omega_n),
\end{equation}
where
$G_s(k,i\omega_n)^{-1}=i\omega_n-E^s_k+\mu-\Sigma(i\omega_n)$ is
the electron propagator in the helicity basis, $\mu$ is the
chemical potential, $\omega_n=(2n+1)\pi T$ is the Matsubara
frequency where $T$ is the temperature, and $\Sigma(i\omega_n)$ is
the impurity self-energy in the self-consistent Born
approximation,
\begin{eqnarray}\label{green2}
\Sigma(i\omega_n)&=&\frac{1}{2\pi\tau
N_0}\int\!\!\frac{d\mathbf{k}}{(2\pi)^2}G(\mathbf{k},i\omega_n)
\nonumber \\
&=&\frac{1}{4\pi\tau N_0
}\sum_{s=\pm}\int\!\!\frac{dk}{2\pi}k\,G_s(k,i\omega_n),
\end{eqnarray}
where $\tau^{-1}=2\pi n_i V_{\rm imp}^2N_0/\hbar$ is the
scattering rate for a 2D electron gas with zero SO interaction and
density of states $N_0=m/2\pi\hbar^2$ per spin direction. The
spin-Hall conductivity is obtained from
\begin{equation}\label{sh1}
\sigma_{\rm sH}=-\lim_{\omega\rightarrow 0}\frac{{\rm
Im}K_{sc}(\omega+i\delta)}{\omega}
\end{equation}
where $K_{sc}$ is the spin-current--charge-current correlation
function given by
\begin{eqnarray}\label{sh2}
K_{sc}(i\omega_m)&=&T\!\sum_n\int\!\!\frac{d\mathbf{k}}{(2\pi)^2}
{\rm Tr}
\left[j^y_s(\mathbf{k})G(\mathbf{k},i\omega_n+i\omega_m)\right.
\nonumber \\
&&\left.J^x_c(\mathbf{k},i\omega_n+i\omega_m,i\omega_n)G(\mathbf{k},i\omega_n)
\right].
\end{eqnarray}
Here $j^y_s(\mathbf{k})=\{S_z,v_y(\mathbf{k})\}/2 =\hbar^2
k_y\sigma_z/2m$ is the current operator in the $y$ direction for
spins polarized along $z$ and $J^x_c$ is the vertex function for
charge current along the $x$ direction. In the Born approximation
for impurity scattering, $J^x_c$ satisfies the following ladder
equation:
\begin{eqnarray}\label{vertex1}
&&J^x_c(\mathbf{k},i\omega_l,i\omega_n)=j^x_c(\mathbf{k})\nonumber
\\ &&\;\;+\frac{1}{2\pi\tau N_0
}\int\!\!\frac{d\mathbf{k}'}{(2\pi)^2}G(\mathbf{k}',i\omega_l)
J^x_c(\mathbf{k}',i\omega_l,i\omega_n)G(\mathbf{k}',i\omega_n),
\nonumber \\
\end{eqnarray}
where $j^x_c(\mathbf{k})=ev_x(\mathbf{k}) =e\hbar k_x/m+ e\gamma
\sigma_y /\hbar$ is the bare charge current. Equation
(\ref{vertex1}) can be rewritten as
$J^x_c(\mathbf{k},i\omega_l,i\omega_n) = e\hbar k_x/m+ e\gamma
\Gamma(i\omega_l,i\omega_n)/\hbar$ where $\Gamma$ represents the
SO corrections to the charge current function. By using
Eq.(\ref{green1}) and by taking advantage of the momentum
independence of $\Sigma$, the correlation function $K_{sc}$
reduces to
\begin{eqnarray}\label{sh3}
K_{sc}(i\omega_m)&=&i\frac{e\hbar^2\gamma}{4m}T\sum_n\Gamma_y(i\omega_n+i\omega_m,i\omega_n)
\nonumber \\
&&\times B_1(i\omega_n+i\omega_m,i\omega_n) \nonumber \\
&\equiv&i\frac{e\hbar^2\gamma}{4m}T\sum_n
\mathcal{K}(i\omega_n+i\omega_m,i\omega_n)
\end{eqnarray}
where $\Gamma_y$ is the component of $\Gamma$ proportional to
$\sigma_y$, $\Gamma_y=\mathrm{Tr}(\sigma_y \Gamma)/2$, which, by
using Eq.(\ref{vertex1}), becomes
\begin{equation}\label{vertex2}
\Gamma_y(i\omega_l,i\omega_n)= \frac{8\pi\tau N_0
+\frac{\displaystyle 1}{\displaystyle k_0}B_2(i\omega_l,i\omega_n)}{8\pi\tau
N_0-B_3(i\omega_l,i\omega_n)}.
\end{equation}
In Eqs.(\ref{sh3}) and (\ref{vertex2}) the function $B_1$, $B_2$,
and $B_3$ are
\begin{eqnarray}
\label{bs1} B_1(i\omega_l,i\omega_n)&=&\int\!\frac{dk}{2\pi}
k^2\!\sum_s\!s\,G_{-s}(k,i\omega_l)G_s(k,i\omega_n),\;\;\;\;\;\; \\
\label{bs2}B_2(i\omega_l,i\omega_n)&=&\int\!\frac{dk}{2\pi}
k^2\!\sum_s\!s\,G_{s}(k,i\omega_l)G_s(k,i\omega_n),\;\;\;\;\;\; \\
\label{bs3}B_3(i\omega_l,i\omega_n)&=&\int\!\frac{dk}{2\pi}
k\!\sum_{s,s'}\,G_{s}(k,i\omega_l)G_{s'}(k,i\omega_n).\;\;\;\;\;\;
\end{eqnarray}

At this point, the analytical continuation to the real axis,
$i\omega_m\rightarrow \omega+i\delta$, can be performed by
following the usual steps,\cite{mahan} leading to
\begin{eqnarray}
\label{sh4} &&T\sum_n \mathcal{K}(i\omega_n+i\omega_m,i\omega_n)
\nonumber \\
&&\!\rightarrow-\int_{-\infty}^{\infty}\!\frac{d\epsilon}{2\pi}
[f(\epsilon+\omega)-f(\epsilon)]{\rm
Im}\mathcal{K}(\epsilon+\omega+i\delta,\epsilon-i\delta)\;\;\;\;\nonumber \\
&&\!-\int_{-\infty}^{\infty}\!\frac{d\epsilon}{2\pi}
[f(\epsilon+\omega)+f(\epsilon)]{\rm Im}
\mathcal{K}(\epsilon+\omega+i\delta,\epsilon+i\delta),
\end{eqnarray}
where $f(x)=1/[\exp(x/T)+1]$ is the Fermi distribution function.
When the spin-Hall conductivity is evaluated via Eq.(\ref{sh1}),
it is clear that the resulting $\sigma_{\rm sH}$ will be given by
the sum of two contributions, $\sigma_{\rm sH}^{RA}$ and
$\sigma_{\rm sH}^{RR}$, respectively, defined as the first and
second line in the right-hand side of Eq.(\ref{sh4}), and
characterized by different combinations of retarded (R) and
advanced (A) Green's functions (see below). The first term,
$\sigma_{\rm sH}^{RA}$, contains in the limit $\omega \rightarrow
0$ the term $df(\epsilon)/d\epsilon$ which, for $T=0$, restricts
all quasiparticle contributions to the Fermi surface. The second
term instead has an integral containing $f(\epsilon)$, allowing
therefore for processes away from the Fermi surface. In
Refs.[\onlinecite{raimo,inoue,olga,loss}] this term has been
disregarded because in the large $E_F$ limit it scales as
$E_0/(\tau E_F^2)$,\cite{raimo,olga} and the spin-Hall
conductivity has been approximated by the $\sigma_{\rm sH}^{RA}$
contribution alone, which at zero temperature reduces to
\begin{equation}
\label{sh5} \sigma_{\rm sH}^{RA}=-\frac{e\hbar^2\gamma}{8\pi
m}\Gamma_y^{RA}\int\!\frac{dk}{2\pi}k^2\sum_s
sG^R_{-s}(k,0)G^A_s(k,0).
\end{equation}
Here $G^{R(A)}_s$ is the retarded (advanced) Green's function and
$\Gamma_y^{RA}=\Gamma_y(0+i\delta,0-i\delta)$ is the ladder
vertex function (\ref{vertex2}) calculated at $i\omega_l=i\delta$
and $i\omega_n=-i\delta$. By assuming that the SO energy $E_0$ is
negligible with respect to $E_F$, then the self-energy
(\ref{green2}) can be approximated as
$\Sigma^R(\omega)=-i/2\tau$,\cite{mahan} and the bubble term $B_2$
defined in Eq.(\ref{bs2}) reduces to
$B_2(i\delta,-i\delta)=-8\pi\tau N_0k_0$, which by using
Eq.(\ref{vertex2}), leads to $\Gamma_y^{RA}=0$. This is the
vertex cancellation mechanism pointed out in
Refs.[\onlinecite{raimo,inoue,olga,loss}].

We reexamine now Eq.(\ref{sh5}) by relaxing the hypothesis
$E_F\gg E_0$. For practical purposes, we introduce an upper
momentum cut-off $k_c$ such that all the relevant momentum and
energy scales are much smaller than the corresponding cut-off
quantities, namely $k_0, k_F \ll k_c$, $E_F, E_0, 1/\tau \ll
E_c=\hbar^2k_c^2/2m$. After the analytical continuation, the
integration over momenta in Eq.(\ref{green2}) can be performed
analytically and the real axis self-energy is evaluated
numerically by iteration. The obtained $\Sigma$ is then
substituted into $\Gamma_y^{RA}$ and $\sigma_{\rm sH}^{RA}$,
Eq.(\ref{sh5}), whose momentum integration allows for an
analytical evaluation due to the momentum independence of
$\Sigma$. To explore the effect of varying $E_F$ on the spin-Hall
conductivity, we have first evaluated the Green's functions at
fixed number electron density $n$, where $n=2$ ($n=0$) means that
all states below the cut-off energy $E_c$ are filled (empty), and
subsequently the corresponding $E_F$ for a given $n$ has been
extracted from
\begin{equation}
\label{density} n=\frac{1}{2E_c}\int_{-\infty}^{\infty}d\omega
f(\omega) \sum_s\frac{N_s(\omega)}{N_0},
\end{equation}
where $N_s(\omega)=-(1/\pi)\int dk/2\pi \, k {\rm Im}
G^R_s(k,\omega)$ is the density of states for the interacting
system and $f(\omega)=\theta(-\omega)$ at zero temperature.

\begin{figure}[t]
\protect
\includegraphics[scale=0.45]{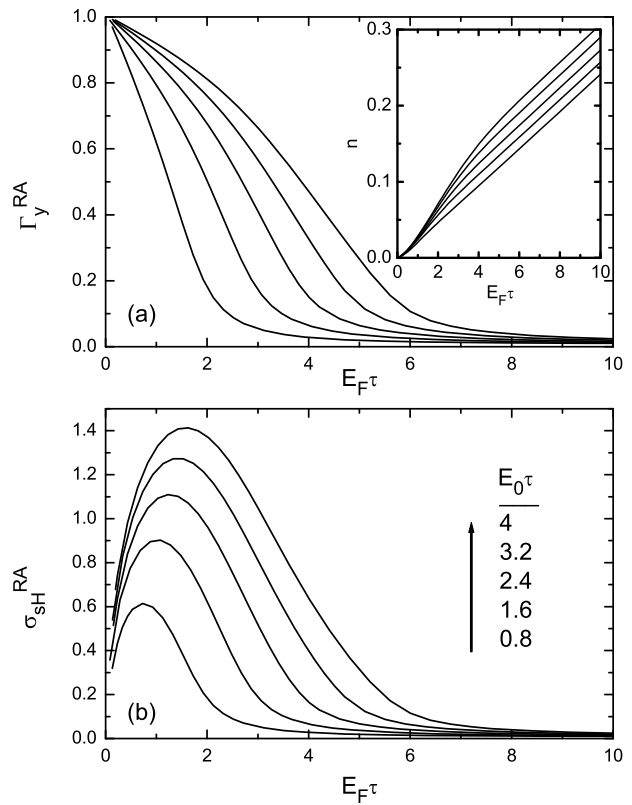}
\caption{(a) Spin-orbit vertex function $\Gamma_y^{RA}$ as a
function of $E_F\tau$ for different values of the spin-orbit
energy $E_0$. The scattering time $\tau$ has been set equal to
$\tau=80/E_c$, where $E_c$ is the upper energy cut-off (see text).
Inset: number electron density $n$ of the interacting system as a
function of $E_F\tau$. (b) The retarded-advanced part $\sigma_{\rm
sH}^{RA}$ of the spin-Hall conductivity in units of $|e|/8\pi$
obtained from Eq.(\ref{sh5}) for the same parameter values of
(a).} \label{fig1}
\end{figure}

In Fig. \ref{fig1}(a) we report the SO vertex function
$\Gamma_y^{RA}$ as a function of $E_F\tau$ and for several values
of the SO energy $E_0$ ranging from $E_0\tau=0.8$ up to
$E_0\tau=4$ (from bottom to top). The coupling to the impurity
potential has been set equal to $E_c\tau=80$ in all cases. The
corresponding values of the number electron density $n$ as a
function of $E_F$ of the interacting system are plotted in the
inset of Fig. \ref{fig1}(a). For $E_F\tau\simeq 10$, the Fermi
energy $E_F$ is sufficiently large compared to $E_0$ and
$\Gamma_y^{RA}$ is negligibly small, confirming the results
reported in Refs.[\onlinecite{raimo,inoue,olga,loss}]. However,
as $E_F\tau$ is decreased, $\Gamma_y^{RA}$ increases
monotonically up to $\Gamma_y^{RA}\simeq 1$ for $E_F/E_0\simeq
0$. In these circumstances, therefore, the vertex cancellation
mechanism is no longer active, and the corresponding spin-Hall
conductivity $\sigma_{\rm sH}^{RA}$ is expected to be non-zero.
This is indeed shown in Fig. \ref{fig1}(b) where $\sigma_{\rm
sH}^{RA}$, Eq.(\ref{sh5}), is plotted in units of $|e|/8\pi$ as a
function of $E_F\tau$. The nonmonotonic behavior of $\sigma_{\rm
sH}^{RA}$ is due to the competition between the increase of
$\Gamma_y^{RA}$ shown in Fig. \ref{fig1}(a) and the decrease of
the integral appearing in Eq.(\ref{sh5}) as $E_F\rightarrow 0$.

A non-vanishing spin-Hall conductivity in an impure 2D Rashba
electron gas is at odds with the general arguments of
Refs.[\onlinecite{olga,rashba}] where $\sigma_{\rm sH}$ has been
shown to be zero for any spin-conserving momentum scattering,
independently of the ratio $E_0/E_F$. However, as already pointed
out above, the physical spin-Hall response is not entirely
defined by Eq.(\ref{sh5}), but should also include the
contributions away from the Fermi surface given by the second
term in the right-hand side of Eq.(\ref{sh4}). Hence $\sigma_{\rm
sH}=\sigma_{\rm sH}^{RA}+\sigma_{\rm sH}^{RR}$, where
\begin{eqnarray}
\label{sh6} \sigma_{\rm sH}^{RR}&=&\frac{e\hbar^2\gamma}{4\pi
m}{\rm Im} \int_{-\infty}^\infty d\epsilon
f(\epsilon)\Gamma_y^{RR}(\epsilon) \nonumber \\
&& \times\int\!\frac{dk}{2\pi}k^2\sum_s
s\frac{dG^R_{-s}(k,\epsilon)}{d\epsilon} G^R_s(k,\epsilon),
\end{eqnarray}
and $\Gamma_y^{RR}(\epsilon)=\Gamma_y(\epsilon+i\delta,\epsilon+i\delta)$.

\begin{figure}[t]
\protect
\includegraphics[scale=0.45]{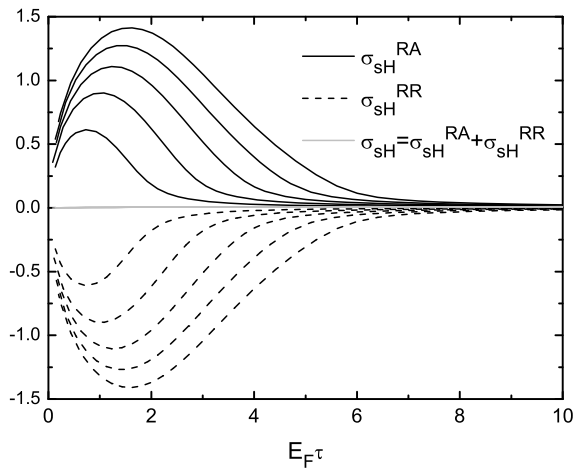}
\caption{The different cotributions to the spin-Hall conductivity
for the same parameters of Fig.\ref{fig1}: $\sigma_{\rm sH}^{RA}$
(solid lines), $\sigma_{\rm sH}^{RR}$ (dashed lines), and the
physical spin-Hall conductivity $\sigma_{\rm sH}=\sigma_{\rm
sH}^{RA}+\sigma_{\rm sH}^{RR}$ (gray lines). All conductivities
are given in units of $|e|/8\pi$.} \label{fig2}
\end{figure}

Our numerical calculations of $\sigma_{\rm sH}^{RR}$,
Eq.(\ref{sh6}), are plotted in Fig. \ref{fig2} (dashed lines)
together with the corresponding $\sigma_{\rm sH}^{RA}$ results
already plotted in Fig. \ref{fig1}(b). For all $E_F/E_0$ values,
$\sigma_{\rm sH}^{RR}$ has the same magnitude of $\sigma_{\rm
sH}^{RA}$ but with opposite sign, so that the resulting physical
spin-Hall conductivity, $\sigma_{\rm sH}=\sigma_{\rm
sH}^{RA}+\sigma_{\rm sH}^{RR}$ (gray lines) reduces to zero within
the accuracy of our numerical calculations.

The results plotted in Fig.\ref{fig2} clearly demonstrate that,
generally, a correct evaluation of the spin-Hall conductivity
must take into account the contributions (\ref{sh6}) away from the
Fermi surface, resolving therefore the concerns expressed in
Ref.[\onlinecite{nomura}] about an only-on-Fermi-surface
cancellation mechanism. However, on this point, a few remarks
should be brought to attention. First, the cancellation between
$\sigma_{\rm sH}^{RA}$ and $\sigma_{\rm sH}^{RR}$ suggests that,
by suitable mathematical transformations, the (nominal) off-Fermi
surface contribution (\ref{sh6}) may be expressed as
$-\sigma_{\rm sH}^{RA}$, resulting in a cancellation mechanism
that is, after all, a Fermi surface property. However, we have
been unable to find such a transformation. A second possibility
is that $\sigma_{\rm sH}^{RR}$ is, generally, a genuine off-Fermi
surface quantity, but that, accidentally, for the model
hamiltonian of Eq.(\ref{ham1}), such a term is quantitatively
equal to $-\sigma_{\rm sH}^{RA}$. In this case, any variation
from the linear Rashba model of (\ref{ham1}) would result in
$\sigma_{\rm sH}^{RA}$ and $\sigma_{\rm sH}^{RR}$ terms which do
not mutually cancel, leading to a nonzero spin-Hall conductivity.
In this respect, one should note that, in fact, the general
arguments of Refs.[\onlinecite{olga,rashba}] about the vanishing
of $\sigma_{\rm sH}$ apply only for model hamiltonians of the
type (\ref{ham1}).

Before concluding, it is worth stressing that the results
presented in this work could be relevant also for systems
described by non-linear Rashba or 3D Dresselhaus SO couplings, or
by non-quadratic unperturbed electronic band structures. It is
known that for such systems, the spin-Hall conductivity in the
presence of momentum scattering is non-zero also for
$E_F/E_0\rightarrow \infty$ because the SO vertex does not
vanish.\cite{mal,mura,nomura,shytov,khaetskii,dassarma} Our
results suggest, however, that even in this case, for finite
$E_F/E_0$, a quantitatively reliable calculation of $\sigma_{\rm
sH}$ should take into account also the off-Fermi surface
contributions. This could be for example the case of the system
studied in Ref.[\onlinecite{koro}] where $E_F$ is of the same
order as $E_0$ and the unperturbed band spectrum is clearly
non-quadratic.

In conclusion, we have calculated the spin-Hall conductivity
$\sigma_{\rm sH}$ for a 2D electron gas subjected to the linear
Rashba SO coupling in the Born approximation for impurity
scattering. We have shown that, apart from the $E_F\rightarrow
\infty$ limit, the spin-dependent part of the vertex function is
nonzero and increases as $E_F\rightarrow 0$, leading to nonzero
Fermi surface contribution $\sigma_{\rm sH}^{RA}$ to the
spin-Hall conductivity. We have demonstrated that the physical
spin-Hall conductivity $\sigma_{\rm sH}$ actually includes also
contributions away from the Fermi surface, $\sigma_{\rm
sH}^{RR}$, which are as large as those on the Fermi surface, but
of opposite sign, leading to a vanishing $\sigma_{\rm sH}$ for
arbitrary values of $E_F/E_0$. We expect that, given the
arguments of Refs.[\onlinecite{olga,rashba}], the mutual
cancellation of $\sigma_{\rm sH}^{RA}$ and $\sigma_{\rm sH}^{RR}$
for $E_F<\infty$ holds true also beyond the self-consistent Born
approximation employed here.

\vspace{10pt}

The authors acknowledge fruitful discussions with Marco Grioni
and Roberto Raimondi. (FM): The hospitality of the Department of
Condensed Matter Physics at the University of Geneva is greatly
appreciated. This work was supported in part by the Natural
Sciences and Engineering Research Council of Canada (NSERC), by
ICORE (Alberta), by the Canadian Institute for Advanced Research
(CIAR), and by the University of Geneva.
%\begin{acknowledgments}
%\end{acknowledgments}

\end{document}